\documentclass[aps,pra,preprint,groupedaddress,superscriptaddress]{revtex4-2} 
\usepackage[colorlinks,linkcolor=red,anchorcolor=blue,citecolor=red]{hyperref} 
\usepackage{braket} 
\usepackage{amsmath} 
\usepackage{amssymb} 
\usepackage[dvips]{graphicx}
\usepackage{color}
\usepackage{verbatim}
\usepackage[T1]{fontenc}
\usepackage{siunitx} 
\usepackage{array} 
\usepackage{multirow} 
\usepackage{booktabs} 
\usepackage{tabularx} 
\usepackage{booktabs} 

\newcolumntype{Y}{>{\centering\arraybackslash}X} 

\makeatletter

\newcommand{\Rmnum}[1]{\expandafter\@slowromancap\romannumeral #1@}
\makeatother 

\begin{document}


\title{Implementing an information-theoretically secure Byzantine agreement with quantum signed message solution}



\author{Yao Zhou}
\author{Feng-Yu Lu}
\affiliation{CAS Key Laboratory of Quantum Information, University of Science and Technology of China, Hefei, Anhui 230026, China}
\affiliation{CAS Center for Excellence in Quantum Information and Quantum Physics, University of Science and Technology of China, Hefei, Anhui 230026, China}

\author{Zhen-Qiang Yin}
\email{yinzq@ustc.edu.cn}
\author{Shuang Wang}
\email{wshuang@ustc.edu.cn}
\author{Wei Chen}
\author{Guang-Can Guo}
\author{Zheng-Fu Han}
\affiliation{CAS Key Laboratory of Quantum Information, University of Science and Technology of China, Hefei, Anhui 230026, China}
\affiliation{CAS Center for Excellence in Quantum Information and Quantum Physics, University of Science and Technology of China, Hefei, Anhui 230026, China}
\affiliation{Hefei National Laboratory, University of Science and Technology of China, Hefei 230088, China}




\begin{abstract}
Byzantine agreement (BA) enables all honest nodes in a decentralized network to reach consensus. In the era of emerging quantum technologies, classical cryptography-based BA protocols face inherent security vulnerabilities. By leveraging the information-theoretic security of keys generated by quantum processing, such as quantum key distribution (QKD), and utilizing the one-time pad (OTP) and one-time universal hashing (OTUH) classical methods proposed in \cite{yin2023QDS}, 
we propose a quantum signed Byzantine agreement (QSBA) protocol based on the quantum signed message (QSM) scheme. This protocol achieves information-theoretic security using only QKD-shared key resources between network nodes, without requiring quantum entanglement or other advanced quantum resources.
Compared to the recently proposed quantum Byzantine agreement (QBA) \cite{weng2023beatingQBA}, our QSBA achieves superior fault tolerance, extending the threshold from nearly 1/2 to an arbitrary number of malicious nodes. Furthermore, our QSBA significantly reduces communication complexity under the same number of malicious nodes.
Simulation results in a 5-node twin-field QKD network highlight the efficiency of our protocol, showcasing its potential for secure and resource-efficient consensus in quantum networks.
\end{abstract}


\maketitle

\section{Introduction}
The Byzantine agreement (BA) \cite{Lamport1982Byzantine} aims to address the challenge of achieving consensus in a distributed system when some nodes may fail or behave maliciously. This agreement is widely used in distributed computing and blockchain technologies \cite{Extance2015Bitcoin}, serving one of their core components.
This consensus problem was abstracted into the Byzantine generals problems by Lamport et al. in 1982 \cite{Lamport1982Byzantine}. In this problem, $n$ ($\ge3$) Byzantine generals each lead an army to besiege a city, with each general located on a different side of the city. The generals can only communicate with each other via messengers. Each general sends their strategy (such as attack or retreat) through his messenger and can also receive strategies from other generals. Eventually, the generals must vote on a common strategy based on their own strategy and the received messages. The challenge arises from the possibility that among the $n$ generals, there may be $m$ traitors, who might send different strategies to the loyal generals to disrupt the consensus decision-making process of the loyal ones. If we focus on analyzing the problem from the perspective of an individual general and his commands, the Byzantine general problem can be translated into the commander-lieutenant model \cite{Lamport1982Byzantine}: A general is randomly selected from the aforementioned $n$ generals to act as the command general, while the remaining $n-1$ generals become lieutenants. The commander general sends his command to $n-1$ lieutenants in such a way that all loyal lieutenants follow the same command. Additionally, if the command general is honest, all loyal lieutenants must execute their command. These two conditions (called the interactive consistency conditions in \cite{Lamport1982Byzantine}) must be satisfied for the original Byzantine generals to reach consensus in this model.

With the advent of the quantum information era, traditional classical cryptography-based Byzantine agreements face potential threats from quantum computing, such as Shor algorithm \cite{Shor1994algorithm} and Grover algorithm \cite{Grover1996algorithm}, which may compromise the core encryption mechanisms of classical Byzantine agreement (CBA). To address the security challenges of CBA, Fitzi et al. first proposed a solution to the three-party consensus using three-qutrit singlet states \cite{Fitzi2001QBA}. This approach marked the first version of a quantum Byzantine agreement (QBA), laying the foundation for its framework and design principles. Since then, numerous researches \cite{Fitzi2002Detectable,Iblisdir2004QBAQKD,Ben2005FastQBA,Gaertner2008DemonstrationQBA,Neigovzen2008CVQBA,Rahaman2015entanglementswapping,Smania2016ExQCP,Kiktenko_2018,Taherkhani_2018,sun2020multi,Wang2022Qblockchain} have introduced various QBA, focusing primarily on improving practicality of these protocols, such as replacing quantum entangled states, and adapting them to broader application scenarios, including extending from three-party to multi-party settings. However, despite these advancements, challenges remain in terms of technical implementation complexity, the completeness of security proofs, and the practicality of consensus message lengths.

In 2023, Weng et al. propose a quantum Byzantine agreement protocol \cite{weng2023beatingQBA} based on the recursive execution of three-party quantum digital signature (QDS) \cite{yin2023QDS,li2023QDS}. By leveraging the key resources generated through quantum processes such as quantum key distribution (QKD) between every node and integrating the one-time pad (OTP) and one-time universal hashing (OTUH) classical methods, this protocol \cite{weng2023beatingQBA} can solve the Byzantine consensus problem in the sense of information-theoretic security without sharing multipartite entangle states. Moreover, it can extend the fault tolerance limit from nearly $1/3$ to $1/2$. Here, we find that in addressing the Byzantine problem, it is not necessary to restrict to three-party QDS. 
The non-repudiation property of QDS is not a necessary condition for solving Byzantine problem. Therefore, we extend the three-party QDS to a multiparty signed message scheme, enabling Byzantine consensus even in the presence of arbitrary malicious nodes.
More importantly, under the fault tolerance limit of nearly $1/2$, our protocol is simpler and consumes fewer resources compared to the proposed QBA in \cite{weng2023beatingQBA}.

The remainder of our paper is organized as follows. In Sec.\ref{QSM-QSBA}, we propose a multiparty quantum signed message (QSM) scheme and use it to form our proposed quantum signed Byzantine agreement (QSBA).
In Sec.\ref{security proof}, we prove the information-theoretic security of both the QSM and QSBA protocols. In Sec.\ref{Communication complexity analysis}, we analyzed the communication complexity of our QSBA protocol and compared it with the QBA protocol in \cite{weng2023beatingQBA}. In a communication network with the same number of malicious nodes, our protocol achieves consensus with significantly lower communication overhead.
In Sec.\ref{simulation}, we demonstrate the signifcant advantages of our QSBA protocol in a five-node network using quantum key resources generated by twin-field quantum key distribution. Finally, we present the summary and conclusion for our paper.
In Sec.\ref{Discussion and Conclusion}, we discuss the reasons why our QSBA significantly reduces communication complexity compared to the QBA protocol and provide a detailed comparative analysis of the basic iterative sub-protocols QDS and QSM of the two protocols. Finally, we summarize the entire paper.

\section{quantum signed message and Byzantine agreement}\label{QSM-QSBA}
The commander-lieutenant model for the Byzantine generals problem is described as: In an n-node network ($n\geq3$), the command general $N_0$ sends their command to $n-1$ lieutenants $N_1$, $N_2$, ..., $N_{n-1}$, subject to the following two consistency conditions:
\begin{enumerate}
	\item[\uppercase\expandafter{\romannumeral1}.] All loyal lieutenants must follow the same command.
	
	\item[\uppercase\expandafter{\romannumeral2}.] If the commander general is loyal, every loyal lieutenant must follow the command issued by the commander.
\end{enumerate}

To address the aforementioned commander-lieutenant model problem, we first introduce the QSM scheme in the $n$-node network, where the nodes are labeled as $N_0$, $N_1$, ..., $N_{n-1}$, and any two nodes $N_i$ and $N_j$ ($i\neq j$) share an information-theoretically secure key $\kappa^{i-j}$ ($=\kappa^{j-i}$). Suppose node $N_j$ publicly announces their message $M$ within the network. The goal is to ensure that any node $N_i$ ($i\neq j$) receiving this message can achieve the following two objectives:

\textbf{\uppercase\expandafter{\romannumeral1}.} Integrity: $N_i$ confirms that the received message $M$ has neither lost bits nor altered by any eavesdropper during transmission.

\textbf{\uppercase\expandafter{\romannumeral2}.} Authentication: $N_i$ can confirm that the message was indeed sent by $N_j$.

Inspired by the OTUH and OTP methods in \cite{yin2023QDS}, we propose the QSM solution for the two objectives described above as follows.

1. For a node $N_i$ ($0\leq i\leq n-1$ and $i\neq j$), $N_j$ randomly choose a hash function $h_{j-i}$ from the $\epsilon$-AXU (almost exclusive-or universal) family of hash functions $H$ and applies the hash function $h_{j-i}$ to the message $M$ to obtain the hash value $h_{j-i}(M)$. Then $N_j$ uses the shared secure key $\kappa^{j-i}$ with $N_i$ to perform OTP encryption on the hash function $h_{j-i}$ and the hash value $h_{j-i}(M)$, resulting in the encrypted hash function $\bar{h}_{j-i}$ and the hash value $\bar{h}_{j-i}(M)$. $N_j$ combines $\bar{h}_{j-i}$ and $\bar{h}_{j-i}(M)$ to form the partial signature value $S_i=\big(\bar{h}_{j-i},\bar{h}_{j-i}(M)\big)$.

2. $N_j$ executes the first step for each other node $N_i$ and combines all the partial signature values to form the final signature matrix value $S^j_{\{0,...,j-1,j+1,...,n-1\}}=(S_0;...;S_{j-1};S_{j+1};...;S_{n-1})$. Finally, $N_j$ combines the message $M$ to be sent with the signature $S^j_{\{0,...,j-1,j+1,...,n-1\}}$ and broadcast the packet $(M:S^j_{\{0,...,j-1,j+1,...,n-1\}})$ to the network.

3. For any node $N_i$ who receives the message packet $(M:S^j_{\{0,...,j-1,j+1,...,n-1\}})$, they can extract the corresponding partial signature value $S_i$ from the signature matrix and obtain the encrypted hash function $\bar{h}_{j-i}$ and the hash value $\bar{h}_{j-i}(M)$. Then $N_i$ can perform the XOR operation using the shared secure key $\kappa^{j-i}$ and get the decrypted hash function $h^\prime_{j-i}$ and the decrypted hash value $h^\prime_{j-i}(M)$. Finally, they apply the decrypted hash function $h^\prime_{j-i}$ to the received message $M$ to obtain the actual hash value $\hat{h}_{j-i}(M)$ and compares it with $h^\prime_{j-i}(M)$. If the two are equal, the signature message sent by $N_j$ is successfully verified by the message receiver; otherwise, the verification fails.

In our QSM scheme, any of the $n-1$ message recipients can verify the message $M$ using $N_j$'s signature $S^j_{\{0,...,j-1,j+1,...,n-1\}}$.
We assume that there are $m$ ($1\leq m\leq n-2$) malicious nodes in the n-node Byzantine network. Any two nodes perform a quantum process, such as QKD, to share an information-theoretically secure keys.
Let $M:S^{i}_{\{i_1,...,i_k\}}$ represents a message signed by node $N_i$ and sent to node group $\{N_{i_1},...,N_{i_k}\}$, and $M:S^{j}_{\{j_1,...,j_k\}}:S^{i}_{\{i_1,...,i_t\}}$ represents a message sequentially signed by nodes $N_j$ and $N_i$, where node $N_j$ sends message to node group $\{N_{j_1},...,N_{j_k}\}$ and node $N_i$ sends message to node group $\{N_{i_1},...,N_{i_t}\}$. Each lieutenant $N_i$ ($1\leq i\leq n-1$) maintains a message set $V_i$, which contain all correctly signed message sequences (rather than the raw message) that they have received. Initially, $V_i$ is an empty set.

\textbf{1. Command issuance:} The command general $N_0$ signs the command $M$ and transmitted the message packet $M:S^{0}_{\{1,...,n-1\}}$ to each lieutenant.

\textbf{2. Message handling and forwarding:}

For each lieutenant $N_i$: 

Upon receiving a signed message in the form $M:S^{0}_{\{1,...,n-1\}}$ directly from $N_0$, and if lieutenant $N_i$ has not already received any message. They initialize their set $V_i$ with the received command, i.e., $V_i=\{M\}$. If $m\geq2$, then they forward the message packet $M:S^{0}_{\{1,...,n-1\}}:S^{i}_{\{1,...,n-1\}-\{i\}}$ to all other lieutenants. Here, we define the subtraction of two sets $G_1$ and $G_2$ as $G_1-G_2=\{x|x\in G_1\ \text{and}\ x\notin G_2\}$.

Upon receiving a signed message in the form $M:S^{0}_{\{1,...,n-1\}}:S^{j_1}_{\{1,...,n-1\}-\{j_1\}}:...:S^{j_k}_{\{1,...,n-1\}-\{j_1\}-...-\{j_{k}\}}$, and if $M$ is not already in their set $V_i$. The message $M$ should be added to their set $V_i$. If the length of the chain of identifiers $k$ is less than a predetermined threshold $m-1$, lieutenant $N_i$ signs the chain and forwards the updated message $M:S^{0}_{\{1,...,n-1\}}:S^{j_1}_{\{1,...,n-1\}-\{j_1\}}:...:S^{j_k}_{\{1,...,n-1\}-\{j_1\}-...-\{j_{k}\}}:S^{i}_{\{1,...,n-1\}-\{j_1\}-...-\{j_{k}\}-\{i\}}$ to all other lieutenants, excluding those whose identifiers are already present in the chain $\{j_1:...:j_k\}$.
If $k=m-1$, lieutenant $N_i$ just forwards the recieved signed message $M:S^{0}_{\{1,...,n-1\}}:S^{j_1}_{\{1,...,n-1\}-\{j_1\}}:...:S^{j_k}_{\{1,...,n-1\}-\{j_1\}-...-\{j_{k}\}}$ to all other lieutenants, excluding $\{N_{j_1}:...:N_{j_k}\}$, through the classical authenticated channel.

\textbf{3. Final decision:}

Once lieutenant $N_i$ has determined that no further messages will be received, they decide on the final command to be executed by applying the decision function $\text{choice}(V_i)$, which generates an order from the message set $V_i$.

\section{security proof } \label{security proof}
We first prove that the QSM scheme ensures integrity and authentication. Building on this, we further demonstrate that the QSBA protocol achieves information-theoretic secure Byzantine consensus.

The information-theoretic security of QSM primarily relies on the security of QKD keys \cite{RN155,doi:10.1126/science.283.5410.2050,PhysRevLett.85.441,RennerPhD,RN127,Hayashi_2012,RevModPhys.94.025008}, the OTP encryption scheme \cite{Shannon1949OTP}, and the property of the $\epsilon$-AXU hash function \cite{Shoup1996divisionhash}. The hash function can map an input message of arbitrary length to a fixed-length hash value of $l_v$-bit. The $\epsilon$-AXU hash function has the collision probability property that for any two distinct message input $m_1\neq m_2$ and any $l_v$-bit string $z$, the probability that a randomly chosen $h\in H$ satisfies $h(m_1)\oplus h(m_2) = z$ is at most $\epsilon$. Note that $\epsilon$-AXU hashing is the generalized version of 2-universal hashing, and $\epsilon\geq \frac{1}{2^{l_v}}$. 

During the QSM process of $N_j$ transmitting a message packet $\big(M,\bar{h}_{j-i},\bar{h}_{j-i}(M)\big)$ to $N_i$, an eavesdropper may intercept the message packet and forge a new one to resend to $N_i$. A successful attack would occur if the forged message packet passed $N_i$'s verification. However, since $\bar{h}_{j-i}$ and $\bar{h}_{j-i}(M)$ are encrypted by OTP with the secure key, the eavesdropper has no knowledge of $h_{j-i}$ and $h_{j-i}(M)$. As a result, the eavesdropper's optimal attack strategy is not to guess $h_{j-i}$ or $h_{j-i}(M)$, but rather to keep $h_{j-i}$ and $h_{j-i}(M)$ unchanged while guessing for a new message $M^\prime\neq M$ such that it produces the same hash value as the original message $M$. Based on the collision probability property of the $\epsilon$-AXU hash function, the probability of the eavesdropper successfully forging the message packet is $\epsilon$. Detailed discussions of this attack can be found in \cite{yin2023QDS} and \cite{li2023QDS}, which align with the conclusion of our simplified analysis. The above security analysis shows that even if the eavesdropper, Eve, possess unlimited computational power, the probability of her successfully forging a message packet that passes verification by the recipient is $\epsilon$.
So in our QSM protocol, if the message sent by $N_j$ is either lost or modified by an eavesdropper, $N_i$ will (most likely) fail the verification. Once $N_i$ successfully verifies the message, it can confirm (with a success probability close to 1) the integrity and authentication of the message.

\begin{figure*}
	\centering
	\includegraphics[width=\textwidth]{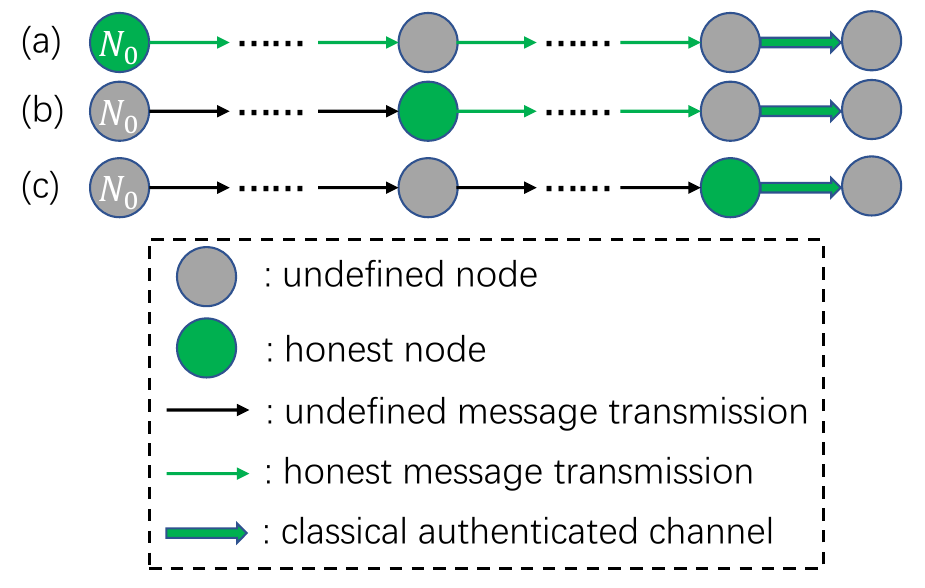}
	\caption{
		The QSBA protocol consists of three types of message propagation chains of length \( m+1 \):  
		(a) When the commander general \( N_0 \) is honest, all honest lieutenants inevitably receive the message sent by \( N_0 \), thereby satisfying consistency condition \uppercase\expandafter{\romannumeral2}.  
		(b) If \( N_0 \) is malicious, honest lieutenants located at positions 2 to \( m \) apply the QSM operation to the received message, ensuring that subsequent honest lieutenants receive a consistent message.  
		(c) An honest lieutenant at the second-to-last position forwards the received message to other honest nodes via a classical authenticated channel. In this case, consistency condition \uppercase\expandafter{\romannumeral1} is evidently satisfied.
	}
	\label{QSBA_message_chain}
\end{figure*}

In the QSBA protocol, the process of reaching consensus can be viewed as consisting of multiple message propagation chains, each of length $m+1$ and involving $m+2$ network nodes. These propagation chains exhibit the property that all honest nodes following an honest node A will must (with a probability close to 1) receive a message consistent with that of A. If node A is the second-to-last node in the message propagation chain, A will transmit a message consistent with the one it received to the final node via a classical authenticated channel. If node A is located at any other position in the chain, it will necessarily apply the QSM to the message received from the previous node. Any malicious node following node A cannot alter or forge A's message and can only forward it truthfully to subsequent nodes. As a result, all honest nodes further along the chain will inevitably receive a message consistent with that of node A. 
In QSBA, the message propagation chains are divided into three categories, (a), (b), and (c), as shown in Fig.\ref{QSBA_message_chain}. If the commander general $N_0$ is honest, all honest lieutenants will inevitably receive the message $M$ sent by $N_0$, satisfying consistency condition \uppercase\expandafter{\romannumeral2}. If the commander general $N_0$ is malicious, in a message propagation chain of length $m+1$, the last node position where a honest lieutenant A appears is $m+1$ (in this case, the first m nodes are all malicious nodes). Lieutenant A can forward the received message to other honest nodes through a classical authentication channel to satisfy consistency condition \uppercase\expandafter{\romannumeral1}. Clearly, the above discussion proves the security of our QSBA protocol.

\section{Communication complexity analysis}\label{Communication complexity analysis}
Our QSBA protocol primarily consists of multiple signed message propagation processes, 
where the signing process involves multiple hash operations, so we can use the number of hash operations to character the communication complexity. For an n-node network with $m$ malicious nodes, the communication complexity $C$ of our QSBA is
\begin{equation}
	C=n-1+(n-1)(n-2)+...+(n-1)\cdot\cdot\cdot(n-m)=\sum_{i=1}^{m} \text{A}_{n-1}^i,
\end{equation}
where $\text{A}_a^b=\frac{a!}{(a-b)!}$ is $b$-permutations of $a$.

In contrast, the QBA proposed in \cite{weng2023beatingQBA} uses the number of QDS executions proposed in \cite{yin2023QDS} to character communication complexity $C^\prime$ as follows:
\begin{equation}
	C^\prime = \sum_{i=2}^{m+1} \text{A}_{n-1}^i.
\end{equation}

The QDS protocol involves a three-party scenario consisting of the message sender Alice, the receiver Bob, and the verifier Charlie.
We assume that Alice and Bob share two 
secure key strings $X_\text{B}$ and $Y_\text{B}$ and Alice and Charlie also share two secure key strings $X_\text{C}$ and $Y_\text{C}$. The protocol flow of QDS is similar to that of QSM.
Alice first randomly chooses a hash function $h$ and applies it to the message $M$ to obtain the hash value $h(M)$. Differently, Alice uses $X_\text{A}\oplus X_\text{B}$ to perform the OTP operation for encrypting the hash function $h$ and $Y_\text{A}\oplus Y_\text{B}$ to perform the OTP operation for encrypting the hash value $h(M)$. Alice then sends the message $M$, along with the encrypted hash function $\bar{h}$ and hash value $\bar{h}(M)$, over an insecure classical channel to Bob. Bob forwards the message packet to Charlie using a classical authenticated channel. Note that, for Bob and Charlie to independently verify the message, they must exchange their secure keys shared with Alice over a classical authenticated channel. This step is essential for them to decrypt the hash function and hash value originally used by Alice.
Note that executing one QDS operation requires the message sender shares a secure key string (the key strings $X$ and $Y$ are combined and treated as a single key string) with each of the receiver and the verifier, performs a hash operation, and two classical authentication channels.
The communication resource consumption comparison of the two protocols are summarized in Tab.\ref{tab:QBA_comparison}.
\begin{table*}[ht]
	\centering
	\caption{A communication resource consumption comparison between two schemes. 
		The original QBA protocol has a communication complexity of $C^\prime$, requiring $C^\prime$ executions of the QDS protocol. Each QDS execution necessitates that the sender and two receivers share a secure string of keys, using the XOR of two key strings to encrypt the hash function and its value, while message transmission relies on two classical authenticated channels. The total communication resources consumed are shown in the second row of the table. 
		Our QSBA protocol has a communication complexity of $C$, requiring $C$ executions of hash operation. The total communication resources consumed are presented in the third row of the table. The last row illustrates the total communication resource savings of our protocol compared to the original protocol.
		Clearly, our scheme saves a significant amount of communication resources and have remarkable advantages.}
	\label{tab:QBA_comparison}
	\begin{tabular}{|c|c|c|c|}
		\hline
		\textbf{} & \textbf{hash operation} & \textbf{secure key string} & \textbf{classical authenticated channel}  \\ \hline
		QBA \cite{weng2023beatingQBA}             & $\sum_{i=2}^{m+1} \text{A}_{n-1}^i$ &  2$\sum_{i=2}^{m+1} \text{A}_{n-1}^i$           & $2\sum_{i=2}^{m+1} \text{A}_{n-1}^i$            \\ \hline
		QSBA             & $\sum_{i=1}^{m} \text{A}_{n-1}^i$ & $\sum_{i=1}^{m} \text{A}_{n-1}^i$            & $\text{A}_{n-1}^{m+1}$            \\ \hline
		advantages             & $\text{A}_{n-1}^{m+1}-\text{A}_{n-1}^{1}$& $\sum_{i=2}^{m+1} \text{A}_{n-1}^i$ + $\text{A}_{n-1}^{m+1}-\text{A}_{n-1}^{1}$           & $\sum_{i=2}^{m+1} \text{A}_{n-1}^i$+$\sum_{i=2}^{m} \text{A}_{n-1}^i$            \\ \hline
	\end{tabular}
\end{table*}

\section{simulation results in a five-node network}\label{simulation}
In this section, we show the significant advantages of our protocol in a 5-node network with at most 2 malicious nodes. In this setup, lieutenants $N_1$, $N_2$, $N_3$ and $N_4$ at different locations form a square, with the commander general $N_0$ located at the center of the square. The distance between $N_0$ and $N_i$ ($1\leq i\leq 4$) is \SI{400}{km}, where the twin-field (TF) QKD protocol is executed, resulting in the sharing of a $\num{2.3895e8}$-bit information-theoretically secure key. An untrusted relay located at $N_0$ is used to implement TFQKD between $N_i$ and $N_j$ ($1\leq i,j\leq 4$ and $i\neq j$), with an actual optical fiber distance of \SI{800}{km}, sharing a 22455-bit information-theoretically secure key. Any two nodes implement the four phase TFQKD with partial phase postselection protocol \cite{Zhou2022PPPTF} and transmit \num{1e14} pulses to achieve \num{4.66e-10}-secure key in the sense of the composable security definition \cite{Renner2009Composable}. Simulation parameters are adopted from \cite{Wang2022TF830}.

Node $N_0$ announces a \SI{0.1}{MB} file, and the other four lieutenants must reach consensus on this file. We adopt the division hash proposed in \cite{Shoup1996divisionhash} to achieve hash operation, where $\epsilon\approx \frac{m}{2^l}$, $m$ denotes the bit length of the message, and $l$ denotes the bit length of the hash value. The hash function maps the \SI{0.1}{MB} file to a 54-bit hash value, achieving a security level of $\num{1e-10}$.

\begin{figure*}
	\centering
	\includegraphics[width=\textwidth]{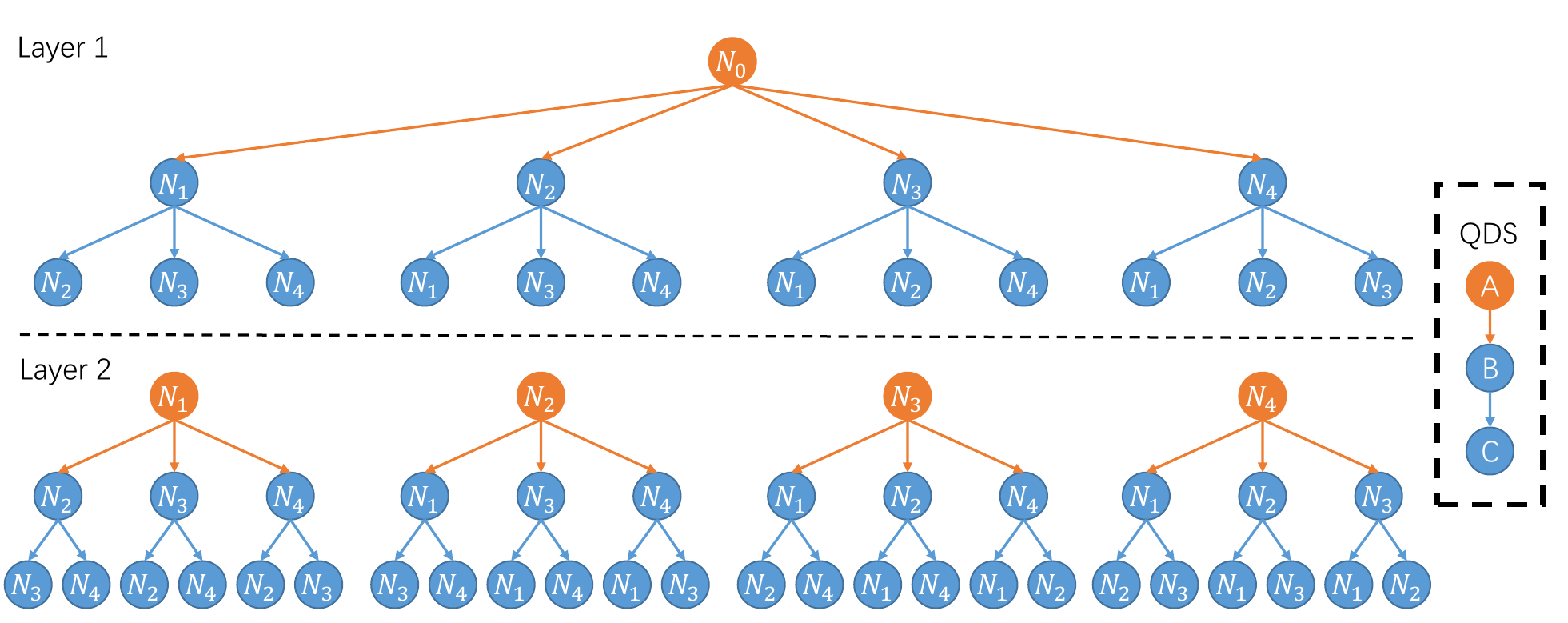}
	\caption{
		The QBA protocol can be regarded as an iterative execution of the QDS protocol. In the first layer, $N_0$ acts as the message sender, while the remaining four nodes sequentially serve as message receivers and verifiers. In the second layer, the QDS protocol is iteratively executed among $N_1$ to $N_4$, each of which acts as the message sender in turn, performing multiple QDS executions similar to those in the first layer.
	}
	\label{QBA_5-node}
\end{figure*}

The detailed message transmission process of the original QBA protocol can be divided into two layers in Fig.\ref{QBA_5-node}. In the first layer, $N_0$ acts as the sender and transmits message to $N_i$ ($i=1,2,3,4$) who then forwards the message to $N_j$ ($j\neq i$), requiring 12 QDS executions. In the second layer, $N_i$ becomes the sender and transmits message to $N_j$ ($j\neq i$) who then forwards the message to $N_k$ ($k\neq i$ and $k \neq j$), requiring 24 QDS executions. A total of 36 QDS executions are performed, enabling lieutenants to determine their final strategies and achieve Byzantine consensus (details in \cite{weng2023beatingQBA}). Here, we focus solely on the communication resources consumed. Key consumption between the commander and each lieutenant is 648 bits, and between any two lieutenants is 864 bits. The protocol performs 36 hash operations and requires 72 instances of classical authenticated channel usage.

\begin{figure*}
	\centering
	\includegraphics[width=\textwidth]{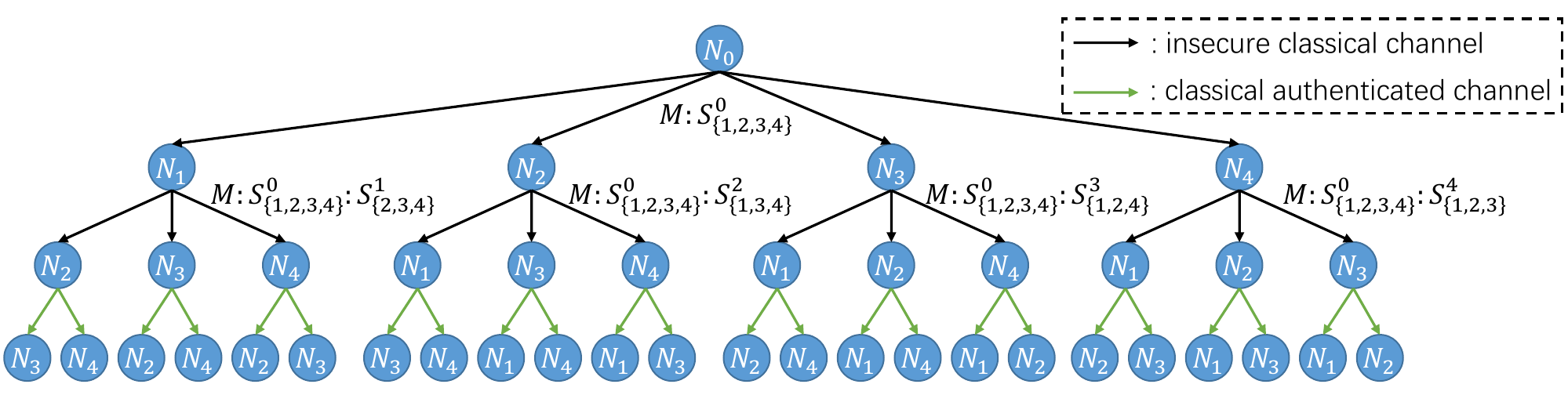}
	\caption{
		In a five-node network with up to two malicious nodes, our QSBA protocol requires only two layers of signatures. The commander general $N_0$ signs the message and sends it to all lieutenants. Each lieutenant then signs the received message and forwards it to the remaining lieutenants. Finally, each lieutenant who receives a doubly signed message only needs to forward it via a classical authenticated channel to the lieutenants who have not yet signed the message.
	}
	\label{QSBA_5-node}
\end{figure*}

\begin{figure*}
	\centering
	\includegraphics[width=\textwidth]{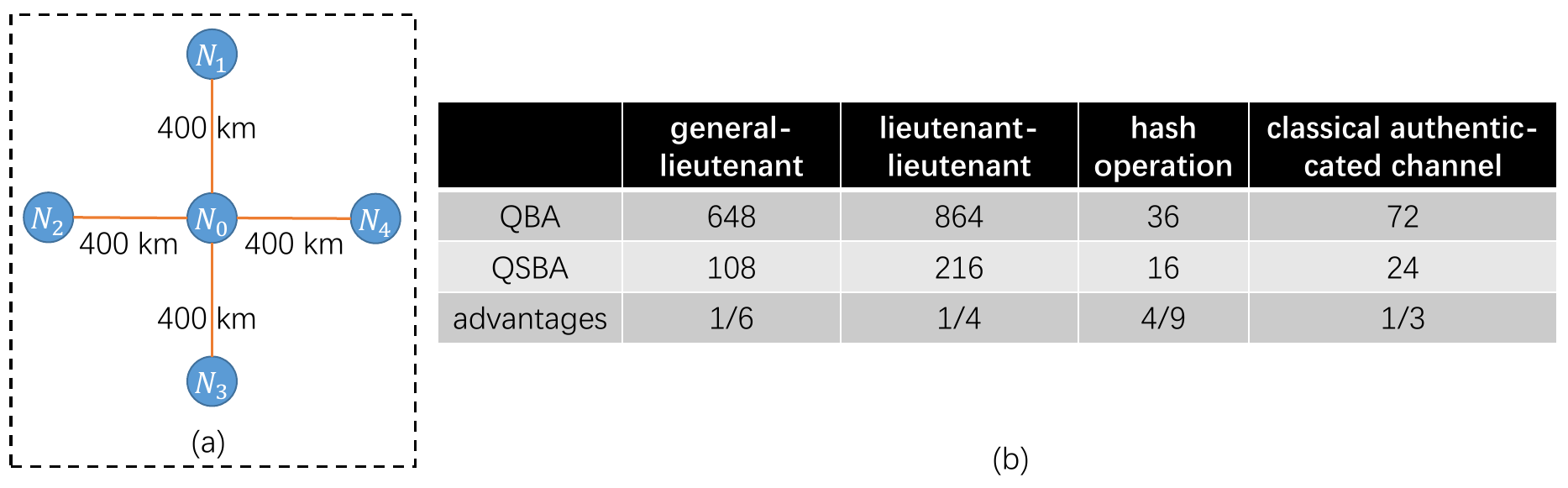}
	\caption{
		In the 5-node network scenario, $N_0$ is the commander general, and $N_1$ to $N_4$ are the lieutenants, with their positional relationships as shown in subfigure (a). Subfigure (b) summarizes the communication resource consumption of the two protocols, with the advantage row indicating the proportion of resource consumption of our protocol compared to the original protocol.
	}
	\label{compare_5-node_result}
\end{figure*}

The detailed message transmission process of our QSBA protocol is shown in Fig.\ref{QSBA_5-node}.
Our protocol requires two layers of signing, with the final layer of message transmission utilizing classical authenticated channels. In the first layer, $N_0$ sends a signed message $M:S^0_{\{1,2,3,4\}}$ to $N_i$ ($i=1,2,3,4$). In the second layer, $N_i$ signs the received message and forward $M:S^0_{\{1,2,3,4\}}:S^i_{\{1,2,3,4\}-\{i\}}$ to $N_j$ ($j\neq i$). Finally, $N_j$ directly sends the received message to $N_k$ ($k\neq i$ and $k \neq j$) using a classical authenticated channel. Key consumption between the commander and each lieutenant is 108 bits, and between any two lieutenants is 216 bits. The protocol performs 16 hash operations and requires 24 instances of classical authenticated channel usage. The comparison of communication resource consumption with the original protocol is shown in Fig.\ref{compare_5-node_result}.

\section{Discussion and Conclusion}\label{Discussion and Conclusion}

Here we discuss the reasons why our QSBA significantly reduces communication complexity compared to the QBA protocol. We find that in the basic iterative sub-protocol QDS of the QBA protocol, its non-repudiation feature is not a necessary condition for achieving Byzantine consensus. In contrast, the basic iterative sub-protocol QSM of our QSBA does not have the non-repudiation feature, thus reducing the communication resources consumed in achieving Byzantine consensus.

Of course, for a complete digital signature scheme, the non-repudiation property is essential. To prevent the message sender from repudiating their message, the message recipient must involve an honest third party as an arbiter to counter such repudiation. To further illustrate the differences between QSM and QDS protocols, we modify the QSM scheme to incorporate non-repudiation properties.
We restrict the QSM scheme to a three-party scenario.
Similar to the QDS three-party scenario mentioned in Sec.\ref{Communication complexity analysis}, we assume that Alice is the message sender, Bob is the message recipient and Charlie is the verifier. Alice and Bob share two secure key strings $X_\text{B}$ and $Y_\text{B}$ and Alice and Charlie also share two secure key strings $X_\text{C}$ and $Y_\text{C}$. In our modified QSM scheme, Alice signs the message $M$ to generate the message package $(M;S)$, where $S=(S_\text{B};S_\text{C})$. Note that $X_\text{B}$ and $Y_\text{B}$ ($X_\text{C}$ and $Y_\text{C}$) combine to constitute the mentioned key $\kappa^{A-B}$ ($\kappa^{A-C}$). Specifically, $X_\text{B}$ ($X_\text{C}$) is used to encrypt the hash function, while $Y_\text{B}$ ($Y_\text{C}$) is used to encrypt the hash value. 
After Bob receives the message packet $(M;S)$ sent by Alice, he verifies the message. If the verification is successful, he forwards the message to Charlie via a classical authenticated channel. Otherwise, Bob directly rejects the message. Charlie then verifies the message packet forwarded by Bob and sends the verification result back to Bob through a classical authenticated channel. Upon receiving Charlie's confirmation of the successful verification, Bob accepts the message; otherwise, he rejects it.

\begin{figure*}
	\centering
	\includegraphics[width=\textwidth]{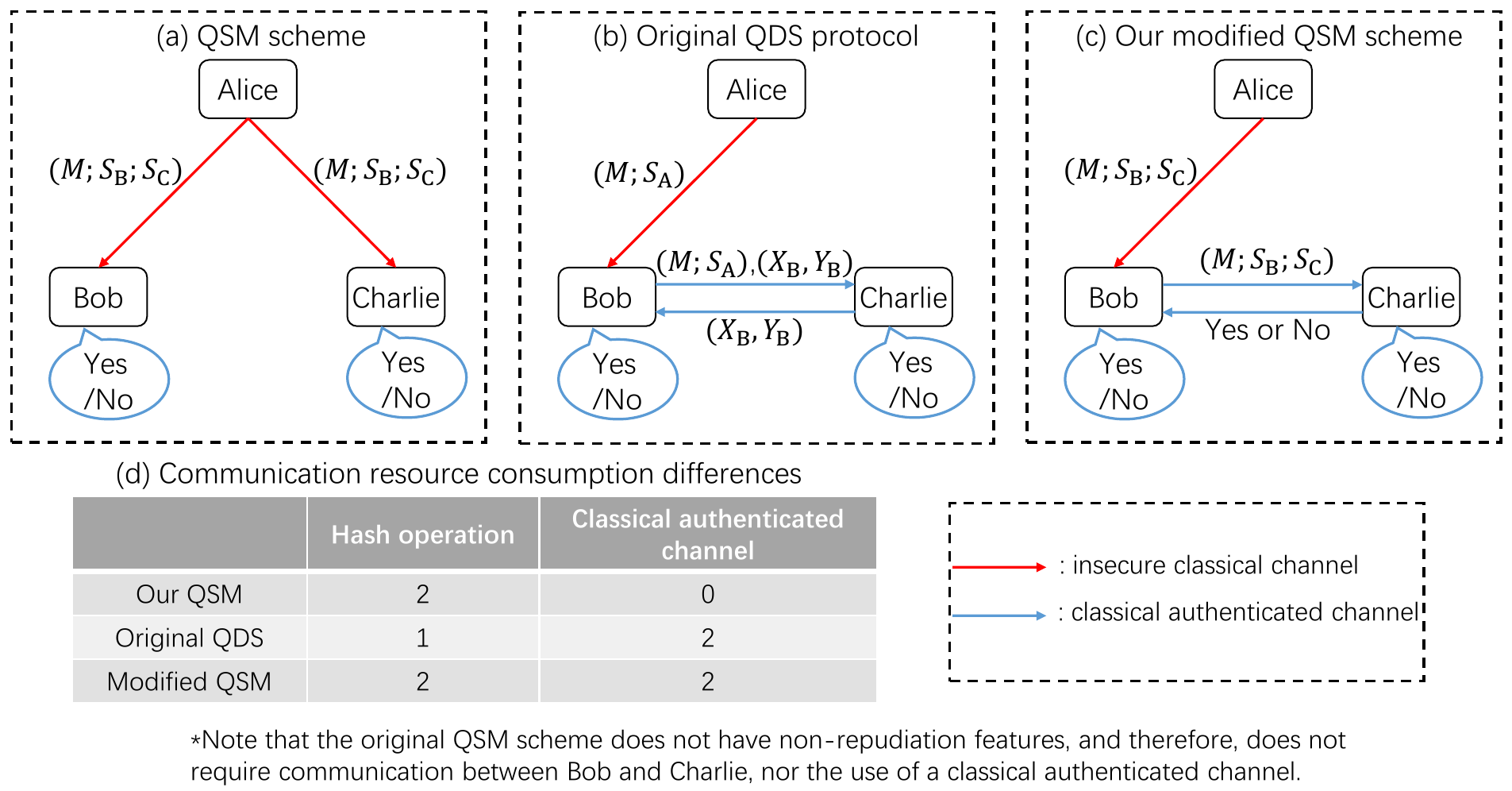}
	\caption{
		Comparison diagram of our (modified) QSM scheme with the original QDS \cite{yin2023QDS}. (a) QSM scheme: Alice signs the message $M$ to generate the packet $(M;S_\text{B};S_\text{C})$ using the shared information-theoretically secure keys with Bob and Charlie and sends it to them over insecure classical channels. Bob and Charlie independently verify the signature and check their results. (b) Original QDS protocol: Alice signs the message $M$ using the keys obtained by performing XOR operations on the two information-theoretically secure key strings that Alice shared with Bob and Charlie. Then She sends the message packet $(M;S_\text{A})$ to Bob over insecure classical channel. Bob forwards the message packet together with the key strings $X_\text{B}$ and $Y_\text{B}$ shared with Alice to Charlie over the classical authenticated channel. Charlie should also transmit the key strings $(X_\text{C},Y_\text{C})$ shared with Alice to Bob over the classical authenticated channel if he passed the verification. Finally, Bob independently verifies the signature and decides whether accepts the message. 
		(c): Modified QSM scheme: Alice generates the message packet $(M;S_\text{B};S_\text{C})$ according to the QSM scheme and sends it to Bob. After Bob successfully verifies the message packet, he forwards it to Charlie. Charlie then sends the verification result to Bob. Bob will only accept the message if Charlie's verification is successful. Note that the communication between Bob and Charlie is carried out through the classical authentication channels.
		(d): the communication resource consumption differences between the three protocols. 
		It is obvious that these protocols follow the communication consumption order: QSM scheme $<$ QDS $<$ modified QSM since construction an information-theoretically secure classical authenticated channel typically requires secure keys and a hash operation.
	}
	\label{QSM-QDS-comparison}
\end{figure*}

The comparison between our (modified) QSM scheme and the original QDS protocol in three-party network is illustrated in Fig.\ref{QSM-QDS-comparison}. For the same message $M$, three protocols require the same consumption of information-theoretically secure keys, and the message recipient need to check whether the verification was successful. Our QSM scheme requires Alice to perform two hash operation but does not rely on classical authenticated channel. The original QDS requires only one hash operation but involves transmitting messages twice over two classical authenticated channels. The modified QSM scheme with non-repudiation property requires two hash operations and utilizes two classical authenticated channels.
We assume that transmitting a message over a classical authenticated channel requires the communication parties to share a string of information-theoretically secure keys and perform one hash operation. Clearly, the communication consumption in the three-party scenario follows the order: QSM $<$ QDS $<$ modified QSM.

In conclusion, we propose an information-theoretically secure QSM scheme which can be modified to form a digital signature protocol with non-repudiation property. Based on the QSM scheme, we present the quantum signed Byzantine agreement. Compared to existing quantum Byzantine agreement protocols, our approach achieves information-theoretic security without relying on quantum entanglement or other advanced quantum resources, utilizing only the shared quantum key resources among nodes. Notably, in contrast to the recently proposed QBA protocol \cite{weng2023beatingQBA}, our method extends the fault tolerance from nearly a 1/2 threshold to arbitrary numbers of malicious nodes. Furthermore, under the same fault-tolerant scenario, our protocol significantly reduces communication resource consumption. Simulation results in a 5-node TFQKD network demonstrate the superiority and efficiency of our proposed protocol.

\begin{acknowledgments}
This work has been supported by the National Key Research and Development Program of China (Grant No. 2020YFA0309802), the National Natural Science Foundation of China (Grant Nos. 62171424, 62271463), Prospect and Key Core Technology Projects of Jiangsu provincial key R \& D Program (BE2022071), the Fundamental Research Funds for the Central Universities, the Innovation Program for Quantum Science and Technology (Grant No. 2021ZD0300701).
\end{acknowledgments}

\bibliography{ref}

\end{document}